\documentclass[copyright,creativecommons]{eptcs}
\providecommand{\event}{CL&C 2010} % Name of the event you are submitting to

\usepackage{amsmath}
\usepackage{amssymb}
\usepackage{mathrsfs}
\usepackage{bussproofs}

\newcommand{\EM}                       { {\tt EM} }
\newcommand{\HA}                       { {\tt HA} }
\newcommand{\Nat}                      { {\tt N} }
\newcommand{\Bool}                     { {\tt Bool} }
\newcommand{\State}                    { {\tt S} }
\newcommand{\NatSet}                   {\mathbb{N}}
\newcommand{\BoolSet}                  {\mathbb{B}}
\newcommand{\StateSet}                 {\mathbb{S}}
\newcommand{\SystemT}                  {\mathsf{T}}
\newcommand{\True}                     { {\tt{True}} }
\newcommand{\False}                    { {\tt{False}} }

\newcommand{\Class}                    {\mbox{\tiny Class}}
\newcommand{\Learn}                    {\mbox{\tiny Learn}}
\newcommand{\SystemTState}             {\SystemT_\State}
\newcommand{\SystemTClass}             {{\SystemT_{\Class}}}
\newcommand{\SystemTLearn}             {{\SystemT_{\Learn}}}

\newcommand{\comment}[1]{}
\newcommand{\proj}                     { {p} }
\newcommand{\CupSem}                   { {\mathcal U} }
\newcommand{\Add}                      { {\mbox{Add}} }
\newcommand{\add}                      { {\mbox{add}} }
\newcommand{\fix}                      { {f} }

\newcommand{\makestate}      [1]       { {\underline{#1}} }
\newcommand{\PRclass}                    {{\mathsf{PCF}_{\Class}}}
\newcommand{\PRlearn}                    {{\mathsf{PCF}_{\Learn}}}

\newtheorem{theorem}{Theorem}

\newtheorem{lemma}{Lemma}
\newtheorem{proposition}{Proposition}
\newtheorem{definition}{Definition}

\title{Interactive Learning Based Realizability and 1-Backtracking Games}
\author{Federico Aschieri
\institute{Dipartimento di Informatica\\
Universit\`a di Torino\\ Italy}
\institute{School of Electronic Engineering and Computer Science\\
Queen Mary, University of London\\
UK}
}

\def\titlerunning{Interactive Learning Based Realizability and 1-Backtracking Games}
\def\authorrunning{F. Aschieri}

\begin{document}
\maketitle

\begin{abstract}  We prove that interactive learning based classical realizability (introduced by Aschieri and Berardi for first order arithmetic \cite{Aschieri}) is sound with respect to Coquand game semantics. In particular, any realizer of an implication-and-negation-free arithmetical formula embodies a winning recursive strategy for the 1-Backtracking version of Tarski games. We also give examples of realizer and winning strategy extraction for some classical proofs. We also sketch some ongoing work about how to extend our notion of realizability in order to obtain completeness with respect to Coquand semantics, when it is restricted to 1-Backtracking games.
 \end{abstract}

\section{Introduction}

In this paper we show that learning based realizability (see Aschieri and Berardi \cite{Aschieri}) relates to 1-Backtracking Tarski games as intuitionistic realizability (see Kleene \cite{Kleene}) relates to Tarski games. It is well know that Tarski games (see, definition \ref{definition-TarskiGame} below) are just a simple way of rephrasing the concept of classical truth in terms of a game between two players - the first one trying to show the truth of a formula, the second its falsehood - and that an intuitionistic realizer gives a winning recursive strategy to the first player. The result is quite expected: since a realizer gives a way of computing all the information about the truth of a formula, the player trying to prove the truth of that formula has a recursive winning strategy. However, not at all \emph{any} classically provable arithmetical formula allows a winning recursive strategy for that player; otherwise, the decidability of the Halting problem would follow. In \cite{Coquand}, Coquand introduced a game semantics for Peano Arithmetic such that, for any provable formula $A$, the first player has a recursive winning strategy, coming from the proof of $A$. The key idea of that remarkable result is to modify Tarski games, allowing players to correct their mistakes and backtrack to a previous position. Here we show that  learning based realizers have direct interpretation as winning recursive strategies in 1-Backtracking Tarski games (which are a particular case of Coquand games see \cite{BerCoq} and definition \ref{definition-1BacktrackingGames} below). The result, again, is expected: interactive learning based realizers, by design, are similar to strategies in games with backtracking: they improve their computational ability by learning from interaction and counterexamples in a convergent way; eventually, they gather enough information about the truth of a formula to win the game. 

An interesting step towards our result was the Hayashi realizability \cite{Hayashi1}. Indeed, a realizer in the sense of Hayashi  represents a recursive winning strategy in 1-Backtracking games. However, from the computational point of view,  realizers do not relate to 1-Backtracking games in a significant way: Hayashi winning strategies work by exhaustive search and, actually, do not learn from the game and from the \emph{interaction} with the other player. As a result of this issue, constructive upper bounds on the length of games cannot be obtained, whereas using our realizability it is possible. For example, in the case of the 1-Backtracking Tarski game for the formula $\exists x \forall y f(x)\leq f(y)$, the Hayashi realizer checks all the natural numbers until an $n$ such that $\forall y f(n)\leq f(y)$ is found; on the contrary, our realizer yields a strategy which bounds the number of backtrackings by $f(0)$, as shown in this paper. In this case, the Hayashi strategy is the same one suggested by the classical \emph{truth} of the formula, but instead one is interested in the constructive strategy suggested by its classical \emph{proof}.

Since learning based realizers are extracted from proofs in $\HA+ \EM_1$ (Heyting Arithmetic with excluded middle over existential sentences, see \cite{Aschieri}), one also has an interpretation of classical proofs as learning strategies. Moreover, studying  learning based realizers in terms of 1-Backtracking games also sheds light on their behaviour and offers an interesting case study in program extraction and interpretation in classical arithmetic. 

The plan of the paper is the following. In section \S \ref{calculusandrealizability}, we recall the calculus of realizers and the main notion of interactive learning based realizability. In section \S \ref{gamesrealizability}, we prove our main theorem: a realizer of an arithmetical formula embodies a winning strategy in its associated 1-Backtracking Tarski game. In section \S \ref{examples}, we extract realizers from two classical proofs and study their behavior as learning strategies. In section \S \ref{completeness}, we define an extension of our realizability and formulate a conjecture about its completeness with respect to 1-Backtracking Tarski games.

\section{The Calculus $\SystemTClass$ and Learning-Based Realizability}\label{calculusandrealizability}

The whole content of this section is based on Aschieri and Berardi \cite{Aschieri}, where the reader may also find full motivations and proofs. We recall here the definitions and the results we need in the rest of the paper.\\
The winning strategies for 1-Backtracking Tarski games will be represented by terms of $\SystemTClass$ (see \cite{Aschieri}). $\SystemTClass$ is a system of typed lambda calculus which extends G\"odel's system $\SystemT$ by adding symbols for non computable functions and a new type $\State$ (denoting a set of states of knowledge) together with two basic operations over it. The terms of $\SystemTClass$ are computed with respect to  a state of knowledge, which represents a finite approximation of the non computable functions used in the system. 

For a complete definition of $\SystemT$ we refer to Girard \cite{Girard}. $\SystemT$ is simply typed $\lambda$-calculus, with atomic types $\Nat$ (representing the set $\NatSet$ of natural numbers) and $\Bool$ (representing the set $\BoolSet = \{\mbox{True},\mbox{False}\}$ of booleans), product types $T \times U$ and arrows types $T \rightarrow U$, constants $0: \Nat$, $\mathsf{S}:\Nat\rightarrow \Nat$, $\True, \False: \Bool$, pairs $\langle.,.\rangle$, projections $\pi_0, \pi_1$, conditional ${\tt if}_T$ and primitive recursion ${\tt R}_T$ in all types, and the usual reduction rules $(\beta),(\pi),({\tt if}),({\tt R})$ for $\lambda$, $\langle .,.\rangle,{\tt if}_T,{\tt R}_T$. From now on, if $t, u$ are terms of $\SystemT$ with $t=u$ we denote provable equality in $\SystemT$. If $k \in \NatSet$, the numeral denoting $k$ is the closed normal term $ S^k(0)$ of type $\Nat$.  All closed normal terms of type $\Nat$ are numerals. Any closed normal term of type $\Bool$ in $\SystemT$ is ${\True}$ or ${\False}$.

We introduce a notation for ternary projections: if $T = A \times (B \times C)$, with $p_0, p_1, p_2$ we respectively denote the terms $\pi_0$, $\lambda x:T.\pi_0(\pi_1(x))$, $\lambda x:T.\pi_1(\pi_1(x))$.
If $u = \langle u_0,\langle u_1,u_2\rangle \rangle : T$, then $\proj_iu=u_i$ in $\SystemT$ for $i=0,1,2$. We abbreviate $\langle u_0,\langle u_1,u_2\rangle \rangle :T $ with $\langle u_0,u_1,u_2\rangle : T$. 

\begin{definition} [States of Knowledge and Consistent Union]\label{definition-StateOfKnowledge}\begin{enumerate}
\item
A $k$-ary {\em predicate} of $\SystemT$ is any closed normal term $P:\Nat^{k}\rightarrow \Bool$ of $\SystemT$.

\item
An atom is any triple $\langle P,\vec{n},{m}\rangle $, where $P$ is a $(k+1)$-ary predicate of $\SystemT$, and $\vec{n},m$ are $(k+1)$ numerals, and $P\vec{n}m = \True$ in $\SystemT$.

\item
Two atoms $\langle P,\vec{n},{m}\rangle $, $\langle P',\vec{n'},{m'}\rangle $ are {\em consistent} if $P = P'$ and $\vec{n} = \vec{n'}$ in $\SystemT$ imply $m = m'$.

\item
A state of knowledge, shortly a {\em state}, is any finite set $S$ of pairwise consistent atoms.

\item
Two states $S_1, S_2$ are consistent if $S_1 \cup S_2$ is a state.

\item
$\StateSet$ is the set of all states of knowledge.
\item
The {\em consistent union} $S_1 \CupSem S_2$ of $S_1, S_2 \in \StateSet$ is $S_1 \cup S_2 \in \StateSet$ minus all atoms of $S_2$ which are inconsistent with some atom of $S_1$.
\end{enumerate}
\end{definition}
For each state of knowledge $S$ we assume having a unique constant $ \makestate{S}$ denoting it; if there is no ambiguity, we just assume that state constants are strings of the form $\{\langle P,\vec{n_1},m_1\rangle,\ldots, \langle P, \vec{n_k}, m_k\rangle\}$, denoting a state of knowledge. We define with $\SystemTState  = \SystemT + \State + \{\makestate{S}|S \in \StateSet\}$ the extension of $\SystemT$ with one atomic type $\State$ denoting $\StateSet$, and a constant $ \makestate{S} : \State$ for each $S \in \StateSet$, and {\em no} new reduction rule. Computation on states will be defined by a set of algebraic reduction rules we call ``functional''.

\begin{definition}[Functional set of rules]\label{definition-functional}
Let $C$ be any set of constants, each one of some type $A_1\rightarrow \ldots \rightarrow A_n\rightarrow A$, for some $A_1,\ldots,A_n, A \in\{ \Bool, \Nat, \State\}$. We say that $\mathcal{R}$ is a {\em functional set of reduction rules} for $C$ if $\mathcal{R}$ consists, for all $c\in C$ and all closed normal terms ${a_1}:A_1,\ldots, {a_n}:A_n$ of $\SystemT_\State$, of exactly one rule $c {a_1}\ldots {a_n}\mapsto {a}$, where ${a}:A$ is a closed normal term of $\SystemT_\State$.
\end{definition}
We define two extensions of $\SystemT_\State$: an extension $\SystemTClass$ with symbols denoting non-computable maps $X_P:\Nat^k\rightarrow \Bool, \Phi_P: \Nat^k\rightarrow \Nat$ (for each $k$-ary predicate $P$ of $\SystemT$) and no computable reduction rules, another extension $\SystemTLearn$, with the computable approximations $\chi_P,\phi_P$ of $X_P, \Phi_P$, and a computable set of reduction rules. $X_P$ and $\Phi_P$ are intended to represent respectively the oracle mapping $\vec{n}$ to the truth value of $\exists x P\vec{n}x$,  and a Skolem function mapping $\vec{n}$ to an element $m$ such that $\exists x P\vec{n}x$ holds iff $P\vec{n}m=\True$.  We use the elements of $\SystemTClass$ to represent non-computable realizers, and the elements of $\SystemTLearn$ to represent a computable ``approximation'' of a realizer. We denote terms of type $\State$ by $\rho, \rho', \ldots$.

\begin{definition} \label{definition-TermLanguageL1}
Assume $P:\Nat^{k+1}\rightarrow \Bool$ is a $k+1$-ary predicate of $\SystemT$. We introduce the following constants:
\begin{enumerate}

\item
$\chi_P:\State \rightarrow \Nat^k\rightarrow \Bool$
and
$\varphi_P:\State \rightarrow \Nat^k\rightarrow \Nat$.

\item
$X_P:\Nat^k\rightarrow \Bool$ and $\Phi_P: \Nat^k \rightarrow \Nat$.

\item
$\Cup:\State\rightarrow \State \rightarrow \State$ (we denote $\Cup\rho_1\rho_2$ with $\rho_1\Cup\rho_2$). 

\item
$\Add_P:\Nat^{k+1} \rightarrow \State$ and $\add_P:\State \rightarrow \Nat^{k+1} \rightarrow \State$.

\end{enumerate}
\begin{enumerate}
\item
$\Xi_\State$ is the set of all constants $\chi_P,\varphi_P, \Cup, \add_P$.

\item
$\Xi$ is the set of all constants $X_P,\Phi_P, \Cup, \Add_P$.

\item
$\SystemTClass = \SystemT_\State + \Xi$.

\item
A term $t \in \SystemTClass$ has state $\makestate{\emptyset}$ if it has no state constant different from $\makestate{\emptyset}$.
\end{enumerate}
\end{definition}
Let $\vec{t} = t_1\ldots t_k$. We interpret $\chi_P{s} \vec{t}$ and $\varphi_P{s}\vec{t} $ respectively as a ``guess'' for the values of the oracle and the Skolem map $X_P$ and $\Phi_P$ for $\exists y.P\vec{t}y$, guess computed w.r.t. the knowledge state denoted by the constant $s$.  There is no set of computable reduction rules for the constants $\Phi_P, X_P \in \Xi$, and therefore no set of computable reduction rules for $\SystemTClass$. If $\rho_1, \rho_2$ denotes the states $S_1, S_2 \in \StateSet$, we  interpret $\rho_1 \Cup \rho_2$ as denoting the consistent union $S_1 \CupSem S_2$ of $S_1, S_2$. $\Add_P$ denotes the map constantly equal to the empty state $\emptyset$. $\add_P{\makestate{S}} \vec{n}m $ denotes the empty state $\emptyset$ if we cannot add the atom $\langle P, \vec{n},m\rangle$ to $S$, either because $\langle P,\vec{n},m'\rangle \in S$ for some numeral $m'$, or because $P\vec{n}m={\False}$. $\add_P{\makestate{S}} \vec{n}m $ denotes the state $\{\langle P, \vec{n},m \rangle\}$ otherwise. We define a system $\SystemTLearn$ with reduction rules over $\Xi_\State$ by a functional reduction set $\mathcal{R}_\State$.

\begin{definition}[The System $\SystemTLearn$] \label{definition-EquationalTheoryL1}
Let  $s, s_1, s_2$ be state constants denoting the states $S, S_1, S_2$. Let $\langle P, \vec{n},m \rangle$ be an atom. $\mathcal{R}_\State$ is the following functional set of reduction rules for $\Xi_\State$:
\begin{enumerate}
\item
If $\langle P,\vec{n},{m}\rangle \in S$, then
$\chi_P{s}\vec{n} \mapsto {\True}$ and $\varphi_P{s}\vec{n} \mapsto {m}$, else
$\chi_P{s}\vec{n} \mapsto {\False}$ and $\varphi_P{s}\vec{n} \mapsto {0}$.

\item
${s_1}\Cup{s_2} \mapsto \makestate{S_1 \CupSem S_2}$
\item
$\add_P{s}\vec{n}{m} \mapsto \makestate{\emptyset}$ if either $\langle P,\vec{n},{m'} \rangle \in S$ for some numeral $m'$ or $P\vec{n}{m} = {\False}$, and $\add_P{s}\vec{n}{m} \mapsto \makestate{\{\langle P,\vec{n},{m} \rangle\}}$ otherwise.

\end{enumerate}

We define $\SystemTLearn = \SystemT_\State + \Xi_\State + \mathcal{R}_\State$.
\end{definition}
\textbf{Remark.} $\SystemTLearn$ is nothing but $\SystemTState$ with some ``syntactic sugar''.  $\SystemTLearn$ is strongly normalizing, has Church-Rosser property for closed term of atomic types and:

\begin{proposition}[Normal Form Property for $\SystemTLearn$]\label{proposition-normalform} Assume $A$ is either an atomic type or a product type. Then any closed normal term $t \in \SystemTLearn$ of type $A$ is: a numeral ${n}:\Nat$, or a boolean $\True,\False:\Bool$, or a state constant $s:\State$, or a pair $\langle u,v \rangle: B \times C$.
\end{proposition}

\begin{definition} Assume $t \in \SystemTClass$ and  $s$ is a state constant. We call ``approximation of $t$ at state $s$'' the term $t[{s}]$ of $\SystemTLearn$ obtained from $t$ by replacing each constant $X_P$ with $\chi_P{s}$, each constant $\Phi_P$ with $\varphi_P{s}$, each constant $\Add_P$ with $\add_P{s}$.
\end{definition}

 If $s, s'$ are state constants denoting $S, S' \in \StateSet$, we write $s \le s'$ for $S \subseteq S'$. We say that a sequence $\{s_i\}_{i\in\NatSet}$ of state constants is a weakly increasing chain of states (is w.i. for short), if $s_i\le s_{i+1}$ for all $i\in\NatSet$.

\begin{definition}[Convergence]
\label{definition-Convergence} Assume
that $\{s_i\}_{i\in\NatSet} $ is a w.i. sequence of state constants,
and $u, v \in \SystemTClass$.
\begin{enumerate}

\item
  $u$ converges in $\{s_i\}_{i\in\NatSet}$ if $\exists i\in\NatSet.
\forall j\geq i.u[s_j]=u[s_{i}]$ in $\SystemTLearn$.

\item
$u$ converges if $u$ converges in every w.i. sequence of state constants.
\end{enumerate}
\end{definition}
Our realizability semantics relies on two properties of the non computable terms of atomic type in $\SystemTClass$. First, if we repeatedly increase the knowledge state $s$, eventually the value of $t[s]$ stops changing. Second, if $t$ has type $\State$, and contains no state constants but $\makestate{\emptyset}$, then we may effectively find a way of increasing the knowledge state $s$ such that eventually we have $t[s]=\makestate{\emptyset}$. 
\begin{theorem}[Stability Theorem] \label{theorem-StabilityTheorem}
Assume $t \in \SystemTClass$ is a closed term of atomic type $A$ ($A\in\{\Bool,\Nat,\State\}$). Then $t$ is convergent.
\end{theorem}

\begin{theorem}[Fixed Point Property]\label{Fixed Point Property}
Let $t:\State$ be a closed term of $\SystemTClass$ of state $\makestate{\emptyset}$, and $s = \makestate{S}$. Define $\tau(S) = S'$ if $t[\makestate{S}] = \makestate{S}'$, and $f(S) = S \cup \tau(S)$.
\begin{enumerate}
\item
For any $n\in\NatSet$, define $f^0(S)=S$ and $f^{n+1}(S)=f(f^n(S))$. There are $h\in\NatSet$, $S'\in\StateSet$ such that ${S}' = \fix^h({S})\supseteq S$, $\fix({S}')={S}'$ and $\tau(S') = \emptyset$.
\item
We may effectively find a state constant $s' \ge s$ such that $t[s'] = \makestate{\emptyset}$.
\end{enumerate}
\end{theorem}

\begin{definition}[The language $\mathcal{L}$ of Peano Arithmetic] \label{definition-extendedarithmetic} 

\begin{enumerate}

\item
The terms of $\mathcal{L}$ are all $t \in \SystemT$, such that $t:\Nat$ and $FV(t) \subseteq \{x_1^\Nat, \ldots, x_n^\Nat\}$ for some $x_1, \ldots, x_n$.

\item
The atomic formulas of $\mathcal{L}$ are all $Qt_1\ldots t_n \in \SystemT$, for some $Q:\Nat^{n}\rightarrow \Bool$ {\em closed term of $\SystemT$}, and some terms $t_1,\ldots,t_n$ of $\mathcal{L}$.

\item
The formulas of $\mathcal{L}$ are built from atomic formulas of $\mathcal{L}$ by the connectives $\lor,\land,\rightarrow \forall,\exists$ as usual.
\end{enumerate}

\end{definition}

\begin{definition}[Types for realizers]
\label{definition-TypesForRealizers} For each
arithmetical formula $A$ we define a type $|A|$ of $\SystemT$ by
induction on $A$:
%\begin{enumerate}
%\item
$|P(t_1,\ldots,t_n)|=\State$,
%\item
$|A\wedge B|=|A|\times |B|$,
%\item
$|A\vee B|= \Bool\times (|A|\times |B|)$,
%\item
$|A\rightarrow B|=|A|\rightarrow |B|$,
%\item
$|\forall x A|=\Nat\rightarrow |A|$,
%\item
$|\exists x A|= \Nat\times |A|$
%\end{enumerate}
\end{definition}
We define now our notion of realizability, which is relativized to a knowledge state $s$, and differs from Kreisel modified realizability for a single detail: if we realize an atomic formula, the atomic formula does not need to be true, unless the realizer is equal to the empty set in $s$.
\begin{definition}[Realizability]
\label{lemma-IndexedRealizabilityAndRealizability}
Assume $s$ is a state constant, $t\in \SystemTClass$ is a closed term of state $\makestate{\emptyset}$, $A \in \mathcal{L} $ is a closed formula, and $t:|A|$. Let $\vec{t} = t_1, \ldots, t_n : \Nat$.

\begin{enumerate}
\item
$t\Vvdash_s P(\vec{t})$ if and only if $t[s]  = \makestate{\emptyset}$ in $\SystemTLearn$ implies
$P(\vec{t})={\True}$

\item
$t\Vvdash_s{A\wedge B}$ if and only if $\pi_0t \Vvdash_s{A}$ and $\pi_1t\Vvdash_s{B}$

\item
$t\Vvdash_s {A\vee B}$  if and only if either $\proj_0t[{s}]={\True}$ in $\SystemTLearn$ and $\proj_1t\Vvdash_s A$, or $\proj_0t[{s}]={\False}$ in $\SystemTLearn$ and $\proj_2t\Vvdash_s B$

\item
$t\Vvdash_s {A\rightarrow B}$ if and only if for all $u$, if $u\Vvdash_s{A}$,
then $tu\Vvdash_s{B}$

\item
$t\Vvdash_s {\forall x A}$ if and only if for all numerals $n$,
$t{n}\Vvdash_s A[{n}/x]$
\item

$t\Vvdash_s \exists x A$ if and only for some numeral $n$, $\pi_0t[{s}]= {n}$ in $\SystemTLearn$ and $\pi_1t \Vvdash_s A[{n}/x]$
\end{enumerate}
We define $t \Vvdash A$ if and only if $t\Vvdash_s A$ for all state constants $s$.
\end{definition}

\begin{theorem}\label{Realizability Theorem}
If $A$ is a closed formula provable in $\HA + \EM_1$ (see \cite{Aschieri}), then there
exists $t\in \SystemTClass$ such that $t\Vvdash A$.
\end{theorem}

\section{Games, Learning and Realizability}\label{gamesrealizability}
\label{section-GamesLearningandRealizability}

In this section, we define the notion of game, its 1-Backtracking version andTarski games. We also prove our main theorem, connecting learning based realizability and 1-Backtracking Tarski games. 

\begin{definition}[Games]
\label{definition-Games}
\begin{enumerate}

 \item
 A \emph{game} $G$ between two players is a quadruple $(V,E_1,E_2, W)$,
where $V$ is a set, $E_1,E_2$ are  subsets of $V\times V$ such that
$Dom(E_1)\cap Dom(E_2)=\emptyset$, where $Dom(E_i)$ is the domain of $E_i$,  and $W$ is a set of sequences,
possibly infinite, of elements of $V$.  The elements of $V$ are
called \emph{positions} of the game; $E_1$, $E_2$ are the transition
relations respectively for player one and player two:
$(v_1,v_2)\in E_i$ means that player $i$ can legally move from the
position $v_1$ to the position $v_2$.

 \item We define a \emph{play} to be a walk, possibly infinite, in the
graph $(V,E_1\cup E_2)$, i.e. a sequence, possibly void,  $v_1::v_2::\ldots:: v_n::\ldots $ of elements of $V$  such that $(v_i, v_{i+1})\in E_1\cup E_2$ for every $i$. A play of the form $v_1:: v_2:: \ldots:: v_n::\ldots $ is said to \emph{start from}  $v_1$. A play is said to be
\emph{complete} if it is either infinite or is equal to $v_1::\ldots:: v_n$ and $v_n\notin Dom(E_1\cup E_2)$. $W$ is required to be a set of
complete plays. If $p$ is a complete play and $p\in  W$, 
%or if $p$is finite and its last element does belong to $Dom(E_2)$, 
we say that player one wins in $p$. If $p$ is a complete play and $p\notin
W$, 
%or if $p$ is finite and its last element does belong to
%$Dom(E_1)$, 
we say that player two wins in $p$.

 \item
 Let $P_G$ be the set of finite plays. Consider a function $f:
P_G\rightarrow V$; a play $v_1::\ldots:: v_n::\ldots$ is said to be
$f$-correct if $f(v_1,\ldots, v_i)=v_{i+1}$ for every $i$ such that
$(v_i,v_{i+1})\in E_1$

 \item
 A \emph{winning strategy} from position $v$ for player one is a function
$\omega: P_G\rightarrow V$ such that every complete
$\omega$-correct play $v::v_1::\ldots :: v_n::\ldots $  belongs to $W$.
 \end{enumerate}
 \end{definition}
 \textbf{Notation.}  If for $i\in \NatSet, i=1,\ldots, n$  we have that $p_i=(p_i)_{0}:: \ldots :: (p_i)_{n_i}$ is a finite sequence of elements of length $n_i$,  with $p_1::\ldots ::p_n$ we denote the sequence \[(p_1)_0::\ldots ::(p_1)_{n_1}:: \ldots :: (p_k)_0::\ldots :: (p_k)_{n_k}\] where $(p_i)_j$ denotes the $j$-th element of the sequence $p_i$. \\\\
 Suppose that $a_1::a_2::\ldots :: a_n$ is a play of a game $G$,
representing, for some reason, a bad situation for player one (for
example, in the game of chess, $a_n$ might be a configuration of
the
chessboard in which player one has just lost his queen). Then,
learnt the lesson, player one might wish to erase some of his moves
and come back to the time the play was just, say, $a_1,a_2$ and
choose, say, $b_1$ in place of $a_3$; in other words, player one
might wish to \emph{backtrack}. Then, the game might go on as
$a_1 :: a_2 ::b_1::\ldots :: b_m$ and, once again, player one might want to
backtrack to, say, $a_1::a_2::b_1::\ldots :: b_i$, with $i< m$, and so
on... As there is no learning without remembering, player one
must keep in mind  the errors made during the play. This is the
idea
of 1-Backtracking games  (for more motivations, we refer the reader to \cite{BerCoq} and \cite{BerLig}) and here is our definition. %\\

\begin{definition}[1-Backtracking Games]
\label{definition-1BacktrackingGames} Let
$G=(V,E_1,E_2,W)$ be a game.

\begin{enumerate}

\item
We define $1Back(G)$ as the game $(P_G, E_1',E_2', W')$, where:\\
\item $P_G$ is the set of finite plays of $G$\\

\item  $E_2':=\{(p::a,\ p::a::b)\ |\  p, p::a\in P_G,
(a,b)\in
E_2 \}$ and \[E_1':=\{(p::a,\ p::a::b)\ |\  p, p::a\in P_G,  (a,b)\in
E_1\}\ \cup\] \[\{(p::a::q::d,\ p::a)\ |\ p, q\in P_G, p::a::q::d\in P_G, a\in Dom(E_1)\]\[
d\notin Dom(E_2), p::a::q::d\notin W \};\]
\item $W'$ is  the set of finite complete plays $p_1::\ldots :: p_n$ of
$(P_G, E_1', E_2')$ such that $p_n\in W$.
\end{enumerate}
\end{definition}
\textbf{Note.} The pair $(p::a::q::d,\ p::a)$ in the definition above of $E_2'$ codifies a {\em backtracking move} by player one (and we point out that $q::d$ might be the empty sequence).\\
\textbf{Remark.} Differently from \cite{BerCoq}, in which both players are allowed to backtrack, we only consider the case in which only player one is supposed do that (as in \cite{Hayashi1}). It is not that our results would not hold: clearly, the proofs in this paper would work just as fine for the definition of 1-Backtracking Tarski games given in \cite{BerCoq}. However, as noted in \cite{BerCoq}, any player-one recursive winning strategy in our version of the game can be effectively transformed into a winning strategy for player one in the other version the game. Hence, adding backtracking for the second player does not increase the computational challenge for player one.
 Moreover, the notion of winner of the game given in \cite{BerCoq} is strictly non constructive and games played by player one with the correct winning strategy may even not terminate. Whereas, with our definition, we can formulate our main theorem as a program termination result: whatever the strategy chosen by player two, the game terminates with the win of player one. This is also the spirit of realizability and hence of this paper: the constructive information must be computed in a finite amount of time, not in the limit. \\

In the well known Tarski games, there are two players and a formula
on the board. The second player - usually called Abelard - tries to
show that the formula is false, while the first player - usually
called Eloise - tries to show that it is true. Let us see the
definition.%\\

\begin{definition}[Tarski Games]
\label{definition-TarskiGame} Let $A$ be a closed
implication and negation free arithmetical formula of $\mathcal{L}$. We define the
Tarski
game for $A$ as the game $T_A=(V, E_1, E_2, W)$, where:
\begin{enumerate}

\item
$V$ is the set of all subformula occurrences of $A$; that is, $V$ is the smallest set of formulas such that, if either $A\lor B$ or $A\land B$ belongs to $V$, then $A,B\in V$; if either $\forall x A(x)$ or $\exists x A(x)$ belongs to $V$, then $A(n)\in V$ for all numerals $n$. \\

\item
$E_1$ is the set of pairs $(A_1,A_2)\in V\times V$ such that   $A_1=\exists x
A(x)$  and $A_2=A(n)$,  or $A_1=A\lor B$ and either $A_2=A$ or
$A_2=B$;\\

\item
$ E_2$ is the set of pairs $(A_1,A_2)\in V\times V$ such that $A_1=\forall x
A(x)$  and $A_2=A(n)$,  or $A_1=A\land B$ and $A_2=A$ or
$A_2=B$;\\

\item
$W$ is the set of finite complete plays $A_1::\ldots ::A_n$ such that
$A_n=\True$.
\end{enumerate}
\end{definition}
\textbf{Note.} We stress that Tarski games are defined only for implication and negation free formulas. Indeed, $1Back(T_A)$, when $A$ contains implications, would be much more involved and less intuitive (for a definition of Tarski games for every arithmetical formula see for example Berardi \cite{Ber2}).\\

What we want to show is that if $t\Vvdash A$,
$t$ gives to player one a recursive winning strategy in
$1Back(T_A)$. The idea of the proof is the following. Suppose we play as player one. Our strategy is relativized to a knowledge state and we start
the game by fixing the actual state of knowledge as $\makestate{\emptyset}$.
Then we play in the same way as we would do in the Tarski game. For
example,  if there is $\forall x A(x)$ on the board and
$A(n)$ is chosen by player two, we recursively play the
strategy given by $tn$; if there is $\exists x A(x)$ on the
board, we calculate $\pi_0t[\makestate{\emptyset}]=n$ and play
$A(n)$ and recursively the strategy given by $\pi_1t$. If there is $A\lor B$ on the board, we calculate $\proj_0t[\makestate{\emptyset}]$, and according as to whether it equals $\True$ or $\False$, we play the strategy recursively given by $\proj_1t$ or $\proj_2t$.
If there is an atomic formula on the board, if it is true, we win; otherwise we extend the current state with the state $\emptyset \Cup t[\makestate{\emptyset}]$, we backtrack and play with respect to the new state of knowledge and trying to keep as close as possible to the previous game.
Eventually, we will reach a state large enough to enable our
realizer to give always correct answers and we will win. Let us consider first an example and then the formal definition of the winning strategy for Eloise.\\

\textbf{Example ($\EM_1$)}. Given a predicate $P$ of $\SystemT$, and its boolean negation predicate $\neg P$ (which is representable in $\SystemT$), the realizer $E_P$ of  \[\EM_1:=\forall x.\ \exists y\
P(x, y)\vee \forall y \neg P(x,y)\]
  is defined as \[\lambda \alpha^{\Nat}
\langle  X_P\alpha,\ \langle
\Phi_P{\alpha},\
\makestate{\emptyset} \rangle ,\ \lambda m^{\Nat}\
\Add_P {\alpha}m\rangle   \]
According to the rules of the game $1Back(T_{\EM_1})$, Abelard is the first to move and, for some numeral $n$, chooses the formula

\[\exists y\ P(n, y)\vee \forall y \neg P(n,y)\]
Now is the turn of Eloise 
and she plays the strategy given by the term

\[\langle  X_Pn,\ \langle
\Phi_P{\alpha},\
\makestate{\emptyset} \rangle ,\ \lambda m^{\Nat}\
\Add_P nm\rangle   \]
Hence, she computes $X_Pn[\makestate{\emptyset}]=\chi_P\makestate{\emptyset} n=\False$ (by definition \ref{definition-EquationalTheoryL1}), so she plays the formula 
\[\forall y \neg P(n,y)\]
and Abelard chooses $m$ and plays 

\[\neg P(n,m)\]
If $\neg P(n,m)=\True$, Eloise wins. Otherwise, she plays the strategy given by 
\[ (\lambda m^{\Nat}\ \Add_P {\alpha}m )m[\makestate{\emptyset}]= \add_P \makestate{\emptyset} n m=\{\langle P,n,m\rangle\}\]
So, the new knowledge state is now $\{\langle P,n,m\rangle\}$ and she backtracks to the formula

\[\exists y\ P(n, y)\vee \forall y \neg P(n,y)\]
Now, by definition \ref{definition-EquationalTheoryL1}, $X_Pn[\{\langle P,n,m\rangle\}]=\True$ and she plays the formula

\[\exists y\ P(n, y)\]
calculates the term \[\pi_0\langle
\Phi_Pn,\
{\emptyset} \rangle[\{\langle P,n,m\rangle\}]=\varphi_P\{\langle P,n,m\rangle\}n=m\]
plays $P(n,m)$ and wins.\\\\
\textbf{Notation.} In the following, we shall denote with upper case letters $A, B,C$
closed arithmetical formulas, with lower case letters $p,q,r$
plays of $T_A$ and with upper case letters $P,Q,R$  plays of
$1Back(T_A)$ (and all those letters may be indexed by numbers). To avoid confusion with the plays of $T_A$, plays of 1Back($T_A$) will be denoted as $p_1,\ldots, p_n$ rather than $p_1::\ldots :: p_n$. Moreover, if $P=q_1,\ldots, q_m$, then $P, p_1,\ldots, p_n$ will denote the sequence $q_1, \ldots, q_m, p_1,\ldots p_n $.

 \begin{definition}\label{adaptrealizer}
 Fix $u$ such that $u\Vvdash A$. Let  $p$ be a finite play of
$T_A$ starting with $A$. We define by induction on the length of
$p$ a term $\rho(p)\in \SystemTClass$ (read as `the realizer adapt
to $p$')  in the following way: \begin{enumerate} 
\item If $p=A$, then
$\rho(p)=u$. \item If $p=(q:: \exists x B(x):: B(n))$ and
$\rho(q:: \exists x B(x))=t$, then $\rho(p)=\pi_1t$. 
\item If $p=(q::
\forall x B(x):: B(n))$ and $\rho(q:: \forall x B(x))=t$,
then $\rho(p)=tn$. 
\item  If $p=(q::  B_0\land B_1:: B_i)$
and $\rho(q:: B_0\land  B_1)=t$, then $\rho(p)=\pi_it$. 
\item If
$p=(q::  B_1\lor B_2:: B_i)$ and $\rho(q:: B_1\lor  B_2)=t$, then
$\rho(p)=\proj_it$.\end{enumerate}
 Given a play $P=Q, q::B$ of $1Back(T_A)$, we set
$\rho(P)=\rho(q::B)$.

\end{definition}

\begin{definition}\label{adaptstate}
 Fix $u$ such that $u\Vvdash A$. Let $\rho$ be as in definition \ref{adaptrealizer} and $P$ be a finite play of $1Back(T_A)$ starting with $A$. We
define by induction on the length of $P$ a state $\Sigma(P)$ (read
as  `the state associated to $P$') in the following way:
\begin{enumerate} 
\item If $P=A$, then $\Sigma(P)=\varnothing$.
\item If $P=(Q, p::B, p::B::C)$ and $\Sigma(Q, p::B)=s$, then
$\Sigma(P)=s$.
 \item If $P=(Q, p::B::q, p::B)$ and $\Sigma(Q,
p::B::q)=s$ and $\rho(Q, p::B::q)=t$, then if $t:\State$, then
$\Sigma(P)=s\Cup t[s]$, else $\Sigma(P)=s$.\end{enumerate}

\end{definition}

\begin{definition}[Winning strategy for 1Back($T_A$)]\label{definition-winningstrategy}
Fix $u$ such that $u\Vvdash A$. Let $\rho$ and $\Sigma$ be respectively as in definition \ref{adaptrealizer} and \ref{adaptstate}. We define a function $\omega$ from the set of finite plays of
$1Back(T_A)$ to set of finite plays of $T_A$; $\omega$ is intended
to be a recursive winning strategy from $A$ for player one in $1Back(T_A)$.
 \begin{enumerate} 
 \item If $\rho(P,q::\exists x B(x))=t$,
$\Sigma(P,q::\exists x B(x))=s$ and $(\pi_0t)[s]={n}$,
then \[\omega(P,q::\exists x B(x))= q::\exists x B(x)::
B({n})\]

 \item If $\rho(P, q::B\lor C)=t$ and $\Sigma(P, q::B\lor C)=s$,
then if $(\proj_0t)[s]=\True$ then \[\omega(P, q::B\lor C)=q::B\lor
C::B\] else \[\omega(P, q::B\lor C)= q::B\lor C:: C\]

 \item If $A_n$ is atomic, $A_n=\False$, $\rho(P, A_1::\cdots::
A_n)=t$ and $\Sigma(P, A_1::\cdots ::A_n)=s$, then \[\omega(P,
A_1::\cdots:: A_n)= A_1::\cdots:: A_i\] where $i$ is equal to the smallest
$j< n$ such that $\rho( A_1::\cdots:: A_j)=w$ and either
\[A_j=\exists x C(x)\land  A_{j+1}= C({n}) \land
(\pi_0w)[s\Cup t[s]]\neq {n}\] 
or   \[A_j=B_1\lor B_2\land
A_{j+1}= B_1 \land (\proj_0w)[s\Cup t[s]]=\False\] 
or \[A_j=B_1\lor B_2\land A_{j+1}= B_2 \land
(\proj_0w)[s\Cup t[s]]=\True\] If such $j$ does not exist, we set $i=n$.
\item
In the other cases, $\omega(P,q)=q$.
\end{enumerate}

\end{definition}

 \begin{lemma} \label{preservationlemma}
 \label{lemma-Completenessof1Backtracking}Suppose $u\Vvdash A$ and $\rho,\Sigma,\omega$ as in definition \ref{definition-winningstrategy}. Let $Q$ be a finite
$\omega$-correct play of $1Back(T_A)$ starting with $A$, $\rho(Q)=t$, $\Sigma(Q)=s$. If
$Q=Q',q'::B$, then $t\Vvdash_s B$. \end{lemma}
\textit{Proof.} By a straightforward induction on the length of
$Q$.

\begin{theorem}[Soundness Theorem] Let $A$ be a closed negation and implication free arithmetical formula. Suppose that $u\Vvdash A$ and consider the game
$1Back(T_A)$. Let $\omega$ be as in definition \ref{definition-winningstrategy}. Then $\omega$ is a recursive winning strategy from $A$ for player one. \end{theorem}
\textit{Proof.} The theorem will be proved in the full version of this paper. The idea is to prove it by contradiction, assuming there is an infinite $\omega$-correct play. Then one can produce an increasing sequence of states. Using theorems \ref{theorem-StabilityTheorem} and \ref{Fixed Point Property}, one can show that Eloise's moves eventually stabilize and that the game results in a winnning position for Eloise.

\section{Examples}
\label{examples}
\comment{
\textbf{Example ($\Sigma^0_1$ and $\Pi^0_2$ formulas).} Suppose
$u\Vdash \exists x P(x)$, with $P(x)$ atomic. Let $n$ be the
smallest natural number such that $s:=(\pi_1u)^n[\varnothing]$ is a
fixed point of $\pi_1u$. We have $u\Vdash_s \exists x P(x)$ and so
$\pi_1u\Vdash_s P(\overline{m})$, where $\overline{m}=(\pi_0u)[s]$.
Since $(\pi_1u)[s]=s$, $P(\overline{m})$ must be true. Hence, here
is the algorithm (in pseudo code) to find the witness for $\exists
x
P(x)$:\\ \\$s:=\varnothing$;\\ repeat $s:=(\pi_1u)[s]$ until
$(\pi_1u)[s]=s$;\\
return $(\pi_0u)[s];$\\ Suppose now $t\Vdash \forall x\exists y
P(x,y)$. Then, given $n\in\NatSet$, $t\overline{n}\Vdash\exists
yP(\overline{n},y)$. So we can apply the algorithm for $\Sigma^0_1$
formulas. Hence, the extracted algorithm is the following:\\
$u:=t\overline{n}$;\\ $s:=\varnothing$;\\ repeat $s:=(\pi_1u)[s]$
until
$(\pi_1u)[s]=s$;\\ return $(\pi_0u)[s];$\\
}

\textbf{Minimum Principle for functions over natural numbers.} The
minimum principle states that every function $f$ over natural
numbers has a minimum value, i.e. there exists an $f(n)\in \NatSet$
such that for every $m\in\NatSet$ $ f(m)\geq f(n)$. We can prove
this principle in $\HA + \EM_1$, for any $f$ in the language. We assume $P(y,x)\equiv f(x)<y$, but, in order to
enhance readability, we will write $f(x)<y$ rather than the obscure
$P(y,x)$. We define:\\ $Lessef(n):= \exists \alpha f(\alpha)\leq
n$\\ $Lessf(n):=\exists \alpha f(\alpha)<n$\\ $Notlessf(n):=
\forall \alpha f(\alpha)\geq n$\\ Then we formulate - in equivalent form - the
minimum principle as: \[Hasminf:=\exists y.\ Notlessf(y)\wedge
Lessef(y)\] The informal argument goes as follows. As base case of the induction, we just observe that $f(k)\leq 0$, implies $f$  has a minimum value (i.e. $f(k)$). Afterwards, if $Notlessf(f(0))$, we are done, we have find the minimum. Otherwise, $Lessf(f(0))$, and hence $f(\alpha)<f(0)$ for some $\alpha$ given by the oracle. Hence $f(\alpha)\leq f(0)-1$ and we conclude that $f$ has a minimum value by induction hypothesis. 

 Now we give the formal proofs, which are natural deduction trees, decorated with terms of $\SystemTClass$, as formalized in \cite{Aschieri}. We first prove
that
$\forall n.\ (Lessef(n)\rightarrow Hasminf)\rightarrow
(Lessef(S(n))\rightarrow Hasminf)$ holds.

\def\proofSkipAmount{\vskip-2ex plus.1ex minus.1ex}
\begin{prooftree}
\small
\AxiomC{$E_P: \forall n.\ Notlessf(S(n))\lor Lessf(S(n))$}
\UnaryInfC{$E_Pn: Notlessf(S(n))\lor Lessf(S(n))$}

                   \AxiomC{$[Notlessf(S(n))]$}
                   \noLine
                   \UnaryInfC{$D_1$}
                   \noLine
                   \UnaryInfC{$Hasminf$}

                                        \AxiomC{$[Lessf(S(n))]$}
                                        \noLine
                                       \UnaryInfC{$D_2$}
                                       \noLine
                                        \UnaryInfC{$Hasminf$}

\TrinaryInfC{$D: Hasminf$}
\UnaryInfC{$\lambda w_2 D: Lessef(S(n))\rightarrow Hasminf$}
\UnaryInfC{$\lambda w_1\lambda w_2 D: (Lessef(n)\rightarrow
Hasminf)\rightarrow  (Lessef(S(n))\rightarrow Hasminf)$}
\UnaryInfC{$\lambda n \lambda w_1\lambda w_2 D: \forall n
(Lessef(n)\rightarrow Hasminf)\rightarrow  (Lessef(S(n)\rightarrow
Hasminf)$}
\end{prooftree}where  the term $D$ is looked at later, $D_1$ is the proof

\def\proofSkipAmount{\vskip0ex plus.1ex minus.1ex}\begin{prooftree}
\small
\AxiomC{$v_1: Notlessf(S(n))$}
                   \AxiomC{$w_2: Lessef(S(n))$}
                   \BinaryInfC{$\langle v_1,w_2\rangle :
Notlessf(S(n))\land
Lessef(S(n))$}
                   \UnaryInfC{$\langle S(n),\langle v_1,w_2\rangle
\rangle : Hasminf$}
\end{prooftree}
and $D_2$ is the proof

\def\proofSkipAmount{\vskip-3ex plus.1ex minus.1ex}
\begin{prooftree}
\small
\AxiomC{$v_2: [Lessf(S(n))]$}
                                        \AxiomC{$w_1:
[Lessef(n)\rightarrow
Hasminf]$}

                                        \AxiomC{$[x_2: f(z)< S(n)]$}
                                        \UnaryInfC{$x_2: f(z)\leq
n$}
                                         \UnaryInfC{$\langle
z,x_2\rangle :
Lessef(n)$}
                                        \BinaryInfC{$w_1\langle
z,x_2\rangle :
Hasminf$}
                                        \BinaryInfC{$w_1\langle
\pi_0v_2,\pi_1v_2\rangle :
Hasminf$}
\end{prooftree}
We prove now that $Lessef(0)\rightarrow Hasminf$
\def\proofSkipAmount{\vskip-4ex plus.1ex minus.1ex}
 \begin{prooftree}
\small
 \AxiomC{$w: [Lessef(0)]$}

 \AxiomC{$x_1: [f(z)\leq 0]$}
 \UnaryInfC{$x_1: f(z)=0$}

 \UnaryInfC{$x_1: f(\alpha)\geq f(z)$}
 \UnaryInfC{$\lambda \alpha x_1: Notlessf(f(z))$}
                \AxiomC{$\emptyset: f(z)\leq f(z)$}
                \UnaryInfC{$\langle z,\emptyset\rangle : Lessef(f(z))$}
 \BinaryInfC{$\langle \lambda\alpha x_1,\langle z,\emptyset\rangle
\rangle : Notlessf(f(z))\land
Lessef(f(z))$}
 \UnaryInfC{$\langle f(z),\langle \lambda\alpha x_1,\langle
z,\emptyset\rangle \rangle \rangle : Hasminf$}
 \BinaryInfC{$\langle f(\pi_0w),\langle \lambda\alpha \pi_1
w,\langle \pi_0w,\emptyset\rangle \rangle \rangle :
Hasminf$}
 \UnaryInfC{$F:=\lambda w\langle f(\pi_0w),\langle \lambda\alpha
\pi_1
w,\langle \pi_0w,\emptyset\rangle \rangle \rangle : Lessef(0)\rightarrow
Hasminf$}
 \end{prooftree}
Therefore we can conclude with the induction rule that \[\lambda
\alpha^\Nat\ {\tt R}F(\lambda n\lambda w_1\lambda w_2 D)\alpha: \forall x.
Lessef(x)\rightarrow Hasminf\] And now the thesis:

\def\proofSkipAmount{\vskip-1ex plus-1ex minus.1ex} \begin{prooftree}
\small
\AxiomC{$\emptyset: f(0)\leq f(0)$}
\UnaryInfC{$\langle 0, \emptyset\rangle : Lessef (f(0))$}
                  \AxiomC{$\lambda
\alpha^\Nat\ {\tt R}F(\lambda n\lambda w_1\lambda w_2 D)\alpha: \forall x. Lessef(x)\rightarrow Hasminf$}
                  \UnaryInfC{${\tt R}F(\lambda n\lambda w_1\lambda w_2
D)f(0): Lessef(f(0))\rightarrow Hasminf$}
\BinaryInfC{$M:={\tt R}F(\lambda n\lambda w_1\lambda w_2 D)f(0)\langle
0,\emptyset\rangle :
Hasminf$}
\end{prooftree}

 Let us now take a closer look to $D$. We have defined \[D:={\tt if}\
X_P S(n)\  {\tt then}\  w_1\langle \Phi_P S(n), \emptyset
\rangle \
{\tt else}\ \langle S(n),\langle \lambda \beta\ (\Add_P) 
S(n)\beta,w_2\rangle \rangle
\]
 Let $s$ be a state and let us consider $M$, the realizer of $Hasminf$, in the base case of the recursion and after in its general form during the computation: ${\tt R}F(\lambda n\lambda
w_1\lambda w_2 D)f(0)\langle m,\emptyset\rangle
[s]$. If $f(0)=0$, \[M[s]={\tt R}F(\lambda n\lambda w_1\lambda
w_2 D)f(0)\langle{0},\emptyset\rangle
[s]=\] \[=F\langle 0,\emptyset\rangle
=\langle f(0),\langle \lambda
\alpha \emptyset, \langle 0,\emptyset\rangle \rangle  \]
If $f(0)=S(n)$, we have two other cases. If $\chi_PsS({n})=\True$, then \[{\tt R}F(\lambda n\lambda w_1\lambda
w_2 D)S({n})\langle{m},\emptyset\rangle
[s]=\]\[=(\lambda n\lambda w_1\lambda w_2 D){n}({\tt R}F(\lambda
n\lambda w_1\lambda w_2 D){n})\langle
{m},\emptyset\rangle [s]= \]\[={\tt R}F(\lambda n\lambda
w_1\lambda w_2 D){n}\langle \Phi_P
(S({n})),\emptyset\rangle [s]\]
If $\chi_PsS({n})=\False$, then \[{\tt R}F(\lambda n\lambda
w_1\lambda w_2 D)S({n})\langle {m},\emptyset\rangle
[s]=\]\[=(\lambda n\lambda w_1\lambda w_2 D){n}({\tt R}F(\lambda
n\lambda w_1\lambda w_2
D){n})\langle {m},\emptyset\rangle [s]= \]\[=
\langle S({n}),\langle \lambda \beta\ (\add_P)s S(n)\beta,\langle {m},\emptyset\rangle \rangle
\rangle \]
In the first case, the minum value of $f$ has been found. In the second case, the operator ${\tt R}$, starting from $S(n)$, recursively calls itself on $n$; in the third case, it reduces to its normal form. From these equations, we easily deduce the behavior of
the realizer of $Hasminf$.   In a pseudo imperative programming
language, for the witness of $Hasminf$ we would write:\\
$n:=f(0);$\\
$while\ (\chi_Psn=\True, i.e.\ \exists m \ such\ that\  f({m})<
n\in
s)$\\
$do\ n:=n-1;$\\
$return\ n;$ \\
Hence, when $f(0)>0$, we have, for some numeral $k$ 
\[M[s]=\langle k,\langle \lambda \beta\ (\add_P)s k\beta,\langle
\varphi_Psk,\emptyset\rangle \rangle \rangle \]
It is clear that $k$  is the
minimum value of $f$, according to the partial information provided by
$s$
about $f$, and that $f(\varphi_Psk)\leq k$. If $s$ is sufficiently complete, then $k$
is the true minimum of $f$.\\
The normal form of the realizer $M$ of $Hasminf$ is so simple that we can immediately extract the winning strategy $\omega$ for the 1-Backtraking version of the Tarski game for $Hasminf$.
Suppose the current state of the game is $s$. If $f(0)=0$, Eloise chooses the formula
\[Notlessf(0)\land Lessef (0)\]
and wins. If $f(0)>0$, she chooses 

\[Notlessf(k)\land Lessef (k)=\forall \alpha\ f(\alpha) \geq k \land \exists \alpha\ f(\alpha)\leq k \]
If Abelard chooses $\exists \alpha\ f(\alpha)\leq k$, she wins, because she responds with $f(\varphi_Psk)\leq k$, which holds. Suppose hence Abelard chooses \[\forall \alpha\ f(\alpha) \geq k\] and then $f(\beta)\geq k$. If it holds, Eloise wins. Otherwise, she adds to the current state $s$
\[(\lambda \beta\ (\add_P)s k\beta)\beta= (\add_P)s k\beta=\{f(\beta)<k\}\]
 and backtracks to $Hasminf$ and then plays again. This time, she chooses 
 
 \[Notlessf(f(\beta))\land Lessef (f(\beta))\]
 (using $f(\beta)$, which was Abelard's counterexample to the minimality of $k$ and is smaller than her previous choice for the minimum value). After at most $f(0)$ backtrackings, she wins. \\

 \textbf{Coquand's Example.} We investigate now an example -  due to
Coquand - in our framework of realizability. We want to prove that
for every function over natural numbers and  for every
$a\in\NatSet$ there exists $x\in\NatSet$ such that $f(x)\leq
f(x+a)$. Thanks to the minimum principle, we can give a very easy
classical proof:
\comment{
\begin{prooftree}
\AxiomC{$h: Hasminf$}
            \AxiomC{$w: Notlessf(\mu)\land Lessef(\mu)$}
            \UnaryInfC{$\pi_1w: Lessef(\mu)$}
                  \AxiomC{$w: Notlessf(\mu)\land Lessef(\mu)$}
                  \UnaryInfC{$\pi_0 w: Notlessf (\mu)$}
                  \UnaryInfC{$\pi_0w(z+a): f(z+a)\geq \mu$}

                                   \AxiomC{$v: f(z)\leq \mu$}
                  \BinaryInfC{$\pi_0w(z+a)\Cup v: f(z)\leq f(z+a)$}
                  \UnaryInfC{$\langle z,\pi_0w(z+a)\Cup v\rangle :
\exists x
f(x)\leq f(x+a)$}
                  \UnaryInfC{$\lambda a \langle z,\pi_0w(z+a)\Cup
v\rangle :
\forall a \exists x f(x)\leq f(x+a)$}
                                    \BinaryInfC{$\lambda a
\langle \pi_0\pi_1w,\pi_0w(\pi_0\pi_1w+a)\Cup
\pi_1\pi_1w\rangle : \forall a \exists
x f(x)\leq f(x+a)$}
\BinaryInfC{$\lambda a
\langle \pi_0\pi_1\pi_1h,\pi_0\pi_1h(\pi_0\pi_1\pi_1h+a)\Cup
\pi_1\pi_1\pi_1h\rangle : \forall a \exists x f(x)\leq f(x+a)$}
\end{prooftree}}

\def\proofSkipAmount{\vskip-1ex plus.1ex minus.1ex}
%\hbox{ \kern-0.2cm
%\scriptsize
\begin{prooftree}
\scriptsize\AxiomC{$Hasminf$}
            \AxiomC{$ [Notlessf(\mu)\land Lessef(\mu)]$}
            \UnaryInfC{$ Lessef(\mu)$}
                  \AxiomC{$ [Notlessf(\mu)\land Lessef(\mu)]$}
                  \UnaryInfC{$ Notlessf (\mu)$}
                  \UnaryInfC{$ f(z+a)\geq \mu$}

                                   \AxiomC{$ [f(z)\leq \mu]$}
                  \BinaryInfC{$ f(z)\leq f(z+a)$}
                  \UnaryInfC{$\exists x
f(x)\leq f(x+a)$}
                  \UnaryInfC{$\forall a \exists x f(x)\leq f(x+a)$}
                                    \BinaryInfC{$ \forall a \exists
x f(x)\leq f(x+a)$}
\BinaryInfC{$\forall a \exists x f(x)\leq f(x+a)$}
\end{prooftree}
%\DisplayProof }
%\end{prooftree}
The extracted realizer is \[\lambda a
\langle \pi_0\pi_1\pi_1M,\pi_0\pi_1M(\pi_0\pi_1\pi_1M+a)\Cup
\pi_1\pi_1\pi_1h\rangle\] where $M$ is the realizer of $Hasminf$.
$m:=\pi_0\pi_1\pi_1M[s]$ is a point the purported minimum value $\mu:=\pi_0M$ of $f$ is attained at, accordingly to the information in the state $s$ (i.e. $f(m)\leq \mu$). So, if Abelard chooses

\[\exists x\ f(x)\leq f(x+a)\]
Eloise chooses

\[f(m)\leq f(m+a)\]
 We have to
consider the term  \[U[s]:=\pi_0\pi_1M(\pi_0\pi_1\pi_1M+a)\Cup
\pi_1\pi_1\pi_1M[s]\] which updates the current state $s$.  Surely,
$\pi_1\pi_1\pi_1M[s]=\emptyset$.  $\pi_0\pi_1M[s]$ is equal either
to $\lambda \beta\ (\add_P)s \mu\beta$
or to $\lambda \alpha \emptyset$.
 So,
what does $U[s]$
actually do? We have:
\[U[s]=\pi_0\pi_1M(\pi_0\pi_1\pi_1M+a)[s]
=\pi_0\pi_1M(m+a)[s]\]
with either $\pi_0\pi_1M(m+a)[s]=\emptyset$ or \[\pi_0\pi_1M(m+a)[s]= \{f(m+a)<f(m)\}\]
So $U[s]$  tests if $f(m+a)< f(m)$; if it is not the
case, Eloise wins, otherwise she enlarges the state $s$, including the information
$f(m+a)<f(m)$ and backtracks to $\exists x f(x)\leq f(x+a)$. Starting from the state $\emptyset$,
after $k+1$ backtrackings, it will be reached a state $s'$, which will
be of the form $\{ f((k+1)a<f(ka),\ldots, f(2a)<f(a), f(a)<f(0)\}$
and Eloise will play $f((k+1)a)\leq f((k+1)a+a)$. Hence, the extracted
algorithm for Eloise's witness is the following:\\\\
$n:=0$; while $f(n)>f(n+a)$ do $n:=n+a$; return $n$;

\section{Partial Recursive Learning Based Realizability and Completeness}\label{completeness}

In this section we extend our notion of realizability and increase the computational power of our realizers, in order to be able to represent any partial recursive function and in particular, we conjecture, every recursive strategies of 1-Backtracking Tarski games. So, we choose to add to our calculus a fixed point combinator $\mathsf{Y}$, such that for every term $u:A\rightarrow A$, $\mathsf{Y}u=u(\mathsf{Yu})$. 

\begin{definition}[Systems $\PRclass$ and $\PRlearn$ ]
We define $\PRclass$ and $\PRlearn$ to be, respectively, the extensions of $\SystemTClass$ and $\SystemTLearn$ obtained by adding for every type $A$ a constant $\mathsf{Y}_A$ of type $(A\rightarrow A)\rightarrow A$ and a new equality axiom $\mathsf{Y_A}u=u(\mathsf{Y_A}u)$ for every term $u: A\rightarrow A$.

\end{definition}

Since in $\PRclass$ there is a schema for unbounded iteration, properties like convergence do not hold anymore (think about a term taking a states $s$ and returning the largest $n$ such that $\chi_Psn=\True$ ). So we have to {\em ask} our realizers to be convergent. Hence, for each type $A$ of $\PRclass$ we define a set $\|A\|$  of
terms $u: A$ which we call the set of {\em stable terms} of
type $A$. We define stable terms by lifting the notion of
convergence from atomic types (having a special case for the atomic
type $\State$, as we said) to arrow and product types.

\begin{definition}[Convergence]
\label{definition-Convergence2} Assume
that $\{s_i\}_{i\in\NatSet} $ is a w.i. sequence of state constants,
and $u, v \in \PRclass$.
\begin{enumerate}

\item
  $u$ converges in $\{s_i\}_{i\in\NatSet}$ if there exists a normal form $v$ such that $\exists i
\forall j\geq i.u[s_j]=v$ in $\PRlearn$.

\item
$u$ converges if $u$ converges in every w.i. sequence of state constants.
\end{enumerate}
\end{definition}

\begin{definition}[Stable Terms]
\label{definition-StableTerms} Let
$\{s_i\}_{i\in\NatSet}$ be a w.i. chain of states and $s \in
\StateSet$. Assume $A$ is a
type. We define a set $\|A\|$ of terms
$t \in\PRclass$ of type $A$, by induction on $A$.
\begin{enumerate}

\item
$\|\State\|=\{t:\State \ |\ t\mbox{ converges}\}$

\item
$\|\Nat\|=\{t:\Nat \ |\ t\mbox{ converges}\}$

\item
$\|\Bool\|=\{t:\Bool \ |\ t\mbox{ converges}\}$

\item
$\|A\times B\|=\{t:A \times B \ |\
\pi_0t\in\|A\|,\pi_1t\in\|B\|\}$

\item
$\|A\rightarrow B\|=\{t:A\rightarrow B \ |\ \forall u\in
\|A\|, tu\in\|B\|\}$
\end{enumerate}

If $t\in\|A\|$, we say that $t$ is a {\em stable} term of type
$A$.
\end{definition}

Now we extend the notion of realizability with respect to $\PRclass$ and $\PRlearn$.

\begin{definition}[Realizability]
\label{lemma-IndexedRealizabilityAndRealizability2}
Assume $s$ is a state constant, $t\in \PRclass$ is a closed term of state $\makestate{\emptyset}$, $A \in \mathcal{L} $ is a closed formula, and $t\in \||A|\|$. Let $\vec{t} = t_1, \ldots, t_n : \Nat$.

\begin{enumerate}
\item
$t\Vvdash_s P(\vec{t})$ if and only if $t[s]  = \makestate{\emptyset}$ in $\PRlearn$ implies
$P(\vec{t})={\True}$

\item
$t\Vvdash_s{A\wedge B}$ if and only if $\pi_0t \Vvdash_s{A}$ and $\pi_1t\Vvdash_s{B}$

\item
$t\Vvdash_s {A\vee B}$  if and only if either $\proj_0t[{s}]={\True}$ in $\PRlearn$ and $\proj_1t\Vvdash_s A$, or $\proj_0t[{s}]={\False}$ in $\PRlearn$ and $\proj_2t\Vvdash_s B$

\item
$t\Vvdash_s {A\rightarrow B}$ if and only if for all $u$, if $u\Vvdash_s{A}$,
then $tu\Vvdash_s{B}$

\item
$t\Vvdash_s {\forall x A}$ if and only if for all numerals $n$,
$t{n}\Vvdash_s A[{n}/x]$
\item

$t\Vvdash_s \exists x A$ if and only for some numeral $n$, $\pi_0t[{s}]= {n}$ in $\PRlearn$ and $\pi_1t \Vvdash_s A[{n}/x]$
\end{enumerate}
We define $t \Vvdash A$ if and only if $t\Vvdash_s A$ for all state constants $s$.
\end{definition}

The following conjecture will be addressed in the next version of this paper:

\begin{theorem}[Conjecture]
Suppose there exists a recursive winning strategy for player one in $1Back(T_A)$. Then there exists a term $t$ of $\PRclass$ such that $t\Vvdash A$.

\end{theorem}

\section{Conclusions and Further work}

The main contribution of this paper is conceptual, rather than technical, and it should be useful to understand the significance and see possible uses of learning based realizability. We have shown how learning based realizers may be understood in terms of backtracking games and that this interpretation offers a way of eliciting constructive information from them. The idea is that playing games represents a way of challenging realizers; they react to the challenge by learning from failure and counterexamples. In the context of games, it is also possible to appreciate the notion of convergence, i.e. the fact that realizers stabilize their behaviour as they increase their knowledge. Indeed, it looks like similar ideas are useful to understand other classical realizabilities (see for example, Miquel \cite{Miq}). 

A further step will be taken in the full version of this paper, where we plan to solve the conjecture about the completeness of learning based realizability with respect to 1Backtracking games. As pointed out by a referee, the conjecture could be interesting with respect to a problem of game semantics, i.e. whether all recursive innocent strategies are intepretation of a term of PCF.


\begin{thebibliography}{99}

\comment{\bibitem{Aschierifull} F.Aschieri, \emph{Interactive Learning Based Realizability and 1Backtracking Games}(Full Version), ( \texttt{http://www.di.unito.it/$\sim$aschieri/1Back.pdf })}
\bibitem{Aschieri} F. Aschieri, S. Berardi,
\emph{Interactive Learning-Based Realizability for Heyting Arithmetic with $\EM_1$}, to appear in Logical Methods in Computer Science, 2010 (preprint: \texttt{http://arxiv.org/abs/1007.1785 })%\end{verbatim}

%\end{verbatim}

\bibitem{Ber2} S. Berardi, \emph{Semantics for Intuitionistic Arithmetic Based on Tarski Games with Retractable Moves.} TLCA 2007

\bibitem{BerLig}S. Berardi, U. De' Liguoro, \emph{Toward the interpretation of non-constructive reasoning as non-monotonic learning}, Information and Computation, vol 207, issue 1, 2009
\bibitem{BerCoq}S. Berardi, T. Coquand, S. Hayashi,
\emph{Games with 1-Bactracking}, to appear in Annals of Pure and Applied Logic, 2010
(see also GALOP 2005). %210-225.



\bibitem{Coquand}T. Coquand,
\emph{A Semantic of Evidence for Classical Arithmetic},
Journal of Symbolic Logic 60, pag 325-337 (1995)


\bibitem{Girard}J.-Y. Girard,
\emph{Proofs and Types},
Cambridge University Press (1989)




\bibitem{Hayashi1}S. Hayashi,
\emph{Can Proofs be Animated by Games?},
Fundamenta Informaticae 77(4), pag 331-343 (2007)


\bibitem{Kleene}S. C. Kleene,
\emph{On the Interpretation of Intuitionistic Number Theory},
Journal of Symbolic Logic 10(4), pag 109-124 (1945)

\bibitem{Miq} A. Miquel, \emph{Relating classical realizability and negative translation for existential witness extraction.} In Typed Lambda Calculi and Applications  (TLCA 2009), pp. 188-202, 2009


\end{thebibliography}
\end{document}